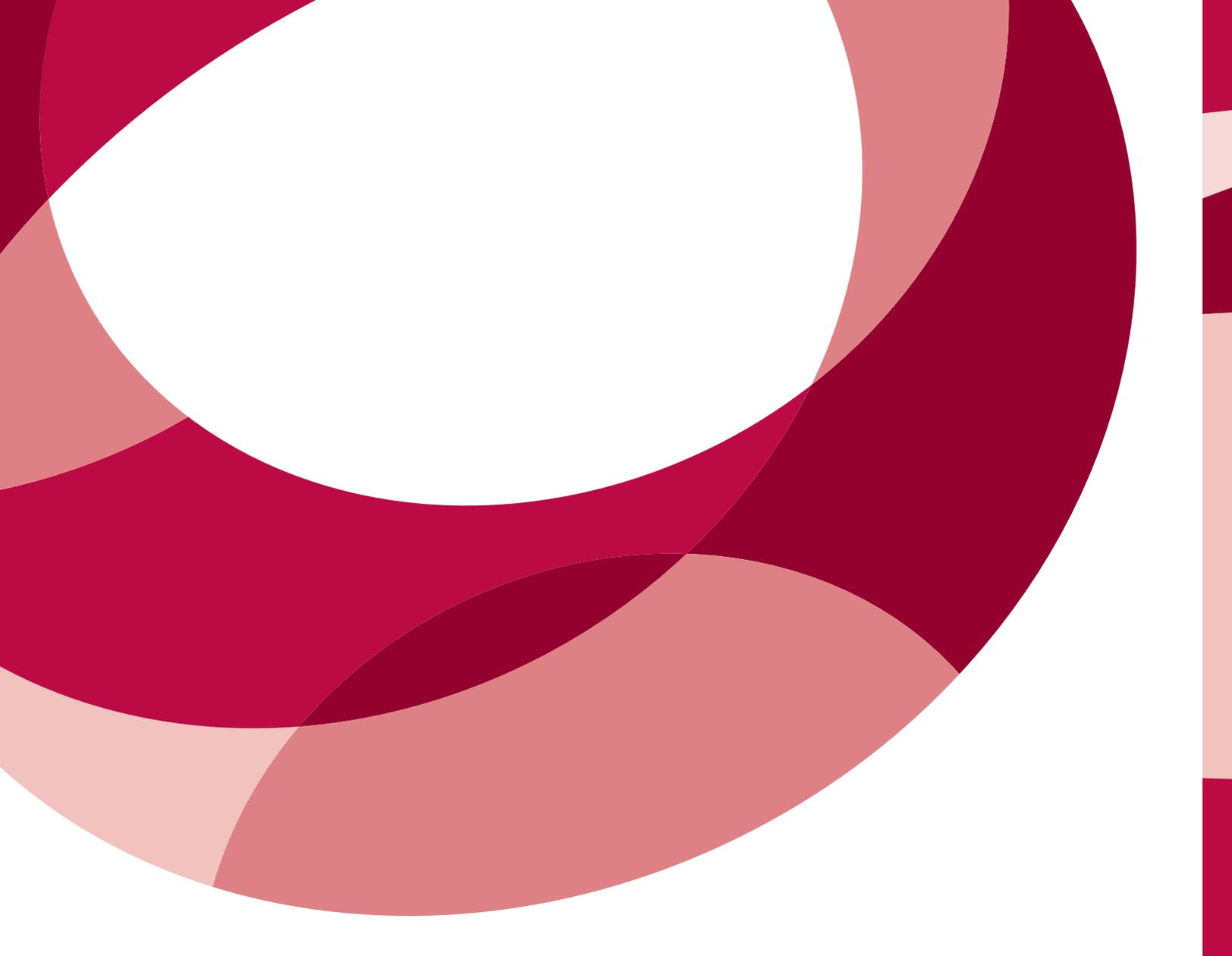

# Future of Pandemic Prevention and Response CCC Workshop Report

February 2024

The material is based upon work supported by the National Science Foundation under Grant No. 1734706. Any opinions, findings, and conclusions or recommendations expressed in this material are those of the authors and do not necessarily reflect the views of the National Science Foundation.

# Future of Pandemic Prevention and Response
# CCC Workshop Report

## Workshop
## September 20-21, 2023

**Workshop Organizers**

David Danks, University of California-San Diego/CCC Council Member*
Rada Mihalcea, University of Michigan/CCC Council Member*
Katie Siek, Indiana University/CCC Council Member*
Mona Singh, Princeton University/CCC Council Member*

**Steering Committee**

Brian E. Dixon, Regenstrief Institute*
Madhav Marathe, University of Virginia
Shwetak Patel, University of Washington
Erica Shenoy, Harvard Medical School and Mass General Brigham
Michael Sjoding, Michigan Medical

**With support from:**

Haley Griffin, Computing Community Consortium*
Catherine Gill, Computing Community Consortium
Ann Schwartz, Computing Community Consortium
Brendan Kane, Computing Research Association

*Report authors

**CCC**
Computing Community Consortium
Catalyst

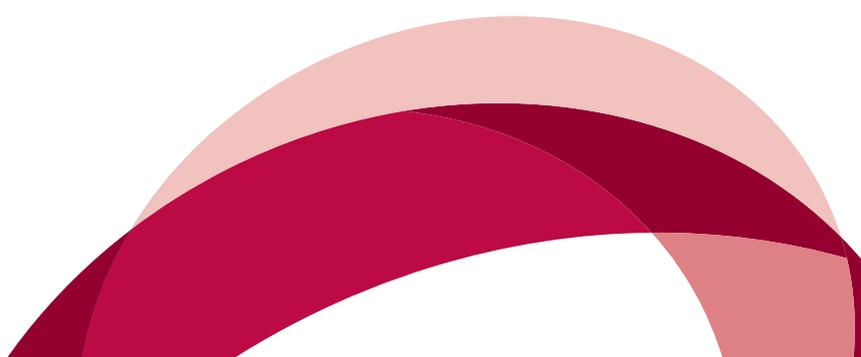

## Executive Summary

This report summarizes the discussions and conclusions of a 2-day multidisciplinary workshop that brought together researchers and practitioners in healthcare, computer science, and social sciences to explore what lessons were learned and what actions, primarily in research, could be taken. One consistent observation was that there is significant merit in thinking not only about pandemic situations, but also about "peacetime" advances, as many healthcare networks and communities are now in a perpetual state of crisis. Attendees discussed how the COVID-19 pandemic amplified gaps in our health and computing systems, and how current and future computing technologies could fill these gaps and improve the trajectory of the next pandemic.

Three major computing themes emerged from the workshop: models, data, and infrastructure. Computational models are extremely important during pandemics, from anticipating supply needs of hospitals, to determining the care capacity of hospital and social service providers, to projecting the spread of the disease. Accurate, reliable models can save lives, and inform community leaders on policy decisions. Health system users require accurate, reliable data to achieve success when applying models. This requires data and measurement standardization across health care organizations, modernizing the data infrastructure, and methods for ensuring data remains private while shared for model development, validation, and application. Finally, many health care systems lack the data, compute, and communication infrastructures required to build models on their data, use those models in ordinary operations, or even to reliably access their data. Although computing research and implementation has shifted heavily towards AI in recent years, older strategies and techniques remain relevant for many real-world challenges today.

A prevalent topic during a wide range of discussions at the workshop was the need for clear and transparent communication with stakeholders. Workshop attendees discussed the need to build trust with community members because the general public will not contribute data, listen to public health guidance influenced by computational models, or use modern technologies without trust and mutual respect between stakeholders. In addition, modernizing the public health infrastructure to capitalize on new software would be a significant investment to ensure all communities can benefit.

## 1. Introduction

The COVID-19 pandemic has taken over 1.1 million American lives[1]. The Computing Community Consortium (CCC) aimed to reflect on how computing was used during the COVID-19 pandemic, how it could have been used more effectively, and what research is needed to improve computing technologies for future responsiveness. CCC Council's "Computational Challenges in Health" Task Force, together with a Steering Committee, brought experts together for a 1.5-day event to see what ideas the health, informatics, epidemiology, Healthcare Personnel (HCP)[2], and computing communities could collectively generate that may mitigate the harm of a future pandemic (see Appendix 2 for a workshop participants list).

The organizers also brought in a team from The Massachusetts General Hospital Center for Disaster Medicine (MGH CDM) to run a Tabletop Exercise on pandemic prevention and response with a focus on computational tools (see Appendix 1). Tabletop exercises are common in the epidemiology/public health world, but it was a new experience for many of the computational experts.

The goals of the workshop were as follows:

◗ Generate creative and dynamic ideas about pandemic prevention and response through brainstorming, collaboration and an emergency simulation tabletop exercise.

◗ Provide space for experts at the intersection of health and computing to network and discuss ideas important to them.

---

[1] https://data.who.int/dashboards/covid19/deaths?m49=840&n=c
[2] Healthcare Personnel definition: https://www.cdc.gov/infectioncontrol/guidelines/healthcare-personnel/terminology.html

## 2. Lessons Learned from the Table Top Exercise

One of the most important lessons that emerged from the tabletop exercise was that it is difficult to build protocols, software, or new policies in times of crisis (i.e., as a pandemic is unfolding). Instead, these processes should be in place and used in non-pandemic times (referred to here informally as "peacetime"). These kinds of peacetime deployments can ensure that processes have been (partially) tested before they are stress-tested and/or further extended during a pandemic. For example, infectious disease modeling can be used in non-pandemic times to track respiratory viruses such as influenza, respiratory syncytial virus (RSV), other communicable diseases, and thus the accuracy of these methods (and their numerous variations) can be continuously evaluated. Similarly, dashboards to perform disease tracking can be made available and routinely used so that their use is normalized prior to a pandemic or other significant health-related event. Overall, computing infrastructure should be developed now so that when a pandemic comes, we know what to do, and we have a strong starting point, rather than starting 'from scratch'. If it is not sensible to use this infrastructure at all times, we need to have set policies for activating these specialized computing systems.

### 2.1. Identifying triggers for pandemics

At the start of an outbreak, one major goal is to figure out if the infection is something that can lead to a pandemic. Genome sequencing of the infectious agent is a top priority, and computational biology approaches comparing the genome to that of existing pathogens can provide significant insight into the (potentially new) infectious agent. Viruses, however, naturally evolve, thus creating a challenge for computational biology approaches to uncover which changes will lead to higher virulence. Research in how to model evolving and not well-defined viruses may better prepare us for the challenges when an outbreak occurs.

We also need to estimate transmission rates, which is challenging due to incomplete knowledge regarding who is infected or the various clinical manifestations of infection. Early in an outbreak or pandemic, before diagnostic tests are widely available, computational approaches analyzing non-traditional sources may be helpful (e.g., google searches, data from wearables, over the counter sales of medicines, movement information from phone or airline travel data, nursing home illnesses, and school absences). Contact tracing will be key, yet this is a time-consuming process—it may be possible to do this more quickly with technological solutions. Computational approaches may also be helpful to determine when to stop contact tracing due to limited usefulness. Multiple different entities (e.g., local public health departments) may be collecting information, thus it is critical to develop data standardization practices. Ideally, researchers and decision makers can use these standardized data in models to help with model training and public health policy decision-making.

### 2.2. Diagnostics

Developing early diagnostic tests will also be critical, and will likely be based on genomic sequence detection. In clinical environments, a major goal is in establishing a case definition, especially before there is a diagnostic test, to better triage incoming patients. Computational approaches may be able to identify trends in infected patients' data to establish case definitions that can be broadly applied by HCP. Even once diagnostic tests are developed, there may not be enough tests available to do widespread testing. It will be important, therefore, to develop computational methods to identify when to perform population screening with integration of appropriate data streams, such as pooled testing and wastewater testing.

It is likely that during a pandemic new computational tools will be developed and specialized for the disease at hand. For example, as we learn more about the illness, it would be helpful to develop computational methods that can predict which patients are at risk of severe illness and/or death. This is especially important if the health system is overwhelmed, so that care can be prioritized for those who need it the most. It is necessary to develop protocols and best practices during peacetime so that newly developed tools and models can be deployed when needed during a pandemic. In addition, use of these tools should be effectively communicated to the broader population so that there are no misunderstandings of prioritization.

## 2.3. Developing treatments

Especially before an effective vaccine becomes available, it will be important to have medical treatments that can stabilize patients and support recovery. The fastest approach will be to try to repurpose existing drugs, but we need better techniques (e.g., computational, experimental) to determine which existing drugs could possibly be repurposed based upon knowledge of the virus and its pathophysiology—it will also be necessary to set up patient trials across healthcare settings.

## 2.4. Infectious disease / epidemiological modeling

Accurate infectious disease modeling would be useful in tracking how a pandemic is unfolding. Currently, numerous such models exist, but their utility in practice has not been benchmarked satisfactorily. While perhaps necessary depending upon what we learn about the infectious agent, it is difficult to *quickly* develop new methods to perform infectious disease modeling as a pandemic is unfolding. Instead, infectious disease modeling can be currently used routinely in clinical practice (as opposed to in academic papers) to track influenza, RSV, covid, other diseases (i.e., in the course of typical public health and healthcare system preparedness and response), and thus the accuracy of these methods (and their numerous variations) can be continuously evaluated.

## 2.5. Supply chain issues

Supply chain issues had and continue to have a major impact in healthcare, especially when travel is restricted, as may be the case during a pandemic. Access to over the counter medications and basic personal hygiene and protective equipment may also be impacted. It is important to have models for how the supply chain works, particularly with respect to medical supplies (including diagnostics, treatments, PPE, cleaning/disinfection supplies, and much more), which can be used to determine how the availabilities of different critical products are (or would be) affected if travel and/or shipping routes are shut down.

## 2.6. Equitable health care

Issues of equity, fairness, and justice in terms of access to medical care and resources will likely be amplified when a pandemic impacts our health care systems. For example, rural communities with poor access to health care as well as other vulnerable populations may be more affected by a pandemic than other groups. Individuals in vulnerable populations[3] may be at increased risk of both exposure and severe disease due to pre-existing conditions, limited access to primary care, and historic prejudice in the health systems[4]. It is necessary to ensure that trial populations are representative when assessing the intended use of drugs, testing, and vaccination against the new disease. Computational models and technologies should avoid bias and explicitly incorporate characteristics of these populations to ensure they are visible and receive appropriate access to care and prevention during a pandemic.

## 2.7. Sharing data resources across medical centers

A major, cross-cutting challenge that emerged from the tabletop simulation is that there are many administrative and logistical challenges to sharing data and resources across various entities, during non-pandemic, peaceful times, and that these are time and labor-prohibitive during a pandemic. Significant barriers include maintaining patient privacy, regulatory laws concerning sharing medical information, limited use of national standards, and the for-profit nature of hospitals not lending itself to cooperation. Additionally, different organizations store their data in myriad ways (e.g., diagnoses may be coded in different ways) and not all health care markets have access to a health information exchange (HIE) network, and this makes data sharing difficult. Therefore, many health systems use local data to train models. Moreover, even these models cannot necessarily be shared due to limited national standards for validating and certifying models. It does not make sense for each health care organization to develop its own unique models and dashboards; these resources should be collaboratively developed and shared. Pooling data and models will enable more robust development and validation of models and dashboards that can better meet community needs during a health crisis.

---

[3] We expand the Office of Priority Populations within the Agency for Healthcare Research and Quality (AHRQ) definition of vulnerable populations that currently includes "delivery of healthcare within inner cities and rural areas; and Healthcare for priority populations, which include: Low income populations, Racial/Ethnic Minorities, Women, Children/Adolescents, Elderly, Individuals with special healthcare needs" https://www.ahrq.gov/priority-populations/about/index.html to also include people with health disparities (e.g., LGBTQI+ people, people with disabilities).

[4] https://arxiv.org/ftp/arxiv/papers/1908/1908.01035.pdf

Given the design of the U.S. health system, it is infeasible to develop a single electronic health record (EHR) system for all organizations. Instead, the health system must leverage systems that can "talk" to each other, or "translate" data between each other. These HIE networks, with support from the Trusted Exchange Framework and Common Agreement (TEFCA) policies called for in the 21st Century Cures Act[5][6], can exchange individual patient data for care as well as population-level data for public health uses. Their use could be expanded to include exchange of case definitions, medical practices, and information on which hospitals have specific resources (e.g., beds or ventilators) available during a pandemic.

## 2.8. Research dissemination

During a pandemic, a wide variety of researchers publish findings intended for a range of audiences and in a range of venues, including preprint servers. The pace of publications and wide range of topics, methods, and outlets presents challenges for interpretation and application. Developing methods to summarize and "rank" this literature is critical. New AI tools to perform meta-analysis of this literature would be particularly helpful.

## 2.9. Science communication

Science communication is critical for ensuring that the population understands community transmission rates, non-pharmaceutical interventions, personal risk, vaccine efficacy, and more. Facilitating information flow through sources that a community trusts is key. Misinformation is a major issue that was discussed during the simulation. Another difficulty is that recommendations can change over time (e.g., mask or don't mask during the recent pandemic), and this may weaken trust in officials despite trial and error, especially in the early days of an outbreak, being a necessary step in the scientific process. Participants discussed that it would be helpful in building public trust to share information about current science knowledge, and healthcare and social service provider capacity, with the public to keep them informed. More work is needed in designing and visualizing uncertain data and communicating the multiple goals, parameters, and interventions to people of varying literacy levels. This information can also be used to decide policy, guide public health recommendations, and healthcare recommendations. AI may also be able to improve messaging; for example, AI could enable rapid, adaptive interactions to help people understand our evolving knowledge about symptom-classes for the disease in times of significant uncertainty (and potentially large amounts of misinformation).

## 3. Paths Forward
### 3.1. Modernizing data infrastructure

One theme that emerged throughout the workshop was the importance of modernizing the data, computing, and communication infrastructures within the healthcare system. On the data side, potentially important information (e.g., patients' recent travel history, levels of PPE and other supplies) is often not recorded. Even when these data are collected, they are frequently difficult to access or use. Electronic health records were noted as a "pain point" in many different contexts, as they can be difficult to use for analysis, modeling, and other efforts to maintain situational awareness about a healthcare network. Even relatively simple queries can be difficult to answer because of the ways that healthcare data are collected and structured. Of course, there are significant privacy concerns about this data; one should not simply collect everything in a single unified data warehouse. Moreover, changes to healthcare data infrastructure must be done with enormous care, as this information must remain stable, accurate, and accessible throughout any updates. Nonetheless, there are significant research opportunities to improve the data infrastructures within healthcare by identifying the (very large) space of common queries, and then adjusting data structures to facilitate answers to those queries. And data collection and use challenges are heightened during a pandemic, so these potential improvements would likely pay dividends during everyday operations.

---

[5] Adler-Milstein, Julia, Chantal Worzala, and Brian E. Dixon. "Chapter 21 - Future Directions for Health Information Exchange." In Health Information Exchange (Second Edition), edited by Brian E. Dixon, 447-68: Academic Press, 2023.

[6] https://www.healthit.gov/topic/interoperability/policy/trusted-exchange-framework-and-common-agreement-tefca

In addition to enhancements to the health care infrastructure, similar investment is needed in the public health data infrastructure. Current Data Modernization Initiative efforts led by the Centers for Disease Control and Prevention (CDC) are necessary but not sufficient to achieve the goals of an operational, robust public health system that can respond to the next pandemic. More efforts are necessary to upgrade technologies, processes, and policies at state and federal levels. Data capture, sharing, and bidirectional communication with the healthcare system is needed. Common approaches across health care and public health should be sought with a goal to enable collaboration and coordination on tackling the nation's toughest health challenges, especially during a pandemic.

## 3.2. Improving computational modeling capabilities

On the computing side, many healthcare networks–in particular, those outside of academic health centers– currently lack the computational capacity and resources to develop, or even run, the types of models that can facilitate situational awareness and improved healthcare delivery. For example, so-called "digital twin" approaches develop computational models of the current state of a system (e.g., a hospital or healthcare network) so that one can do real-time planning, prediction, and response. That is, one builds a digital "twin" that mirrors the current real-world system, precisely since the digital twin can be used to predict, evaluate alternative policies, and so on for significantly less cost (in time and money). However, there are few healthcare systems that have ready access to the level of computing resources that would be required to run such digital twins. More generally, models of disease spread within a community, or rapid search for personalized treatments, or a host of other biomedical advances all require significant compute resources that are largely absent from healthcare systems at the moment. Cloud computing providers could potentially fill this gap, but the security and privacy needs of healthcare systems would likely lead to significant price increases for this type of on-demand cloud computing, thereby potentially putting these resources out of reach. There is thus a significant need to increase the compute capabilities of healthcare networks, including not only the computing devices, but also the human capabilities to manage those resources. Alternately, healthcare systems could take better advantage of their existing computational resources (e.g., computers at nursing stations or in administrative offices), particularly those distributed throughout the systems for compute power. However, this approach would require significant research into algorithms for running computing-intensive models (e.g., digital twins) in highly distributed environments with unpredictable availability.

## 3.3. Enhancing communication infrastructure

A consistent theme of the workshop was the lack of robust communication infrastructure for healthcare data. Most notably, significant volumes of healthcare information (e.g., diagnostic testing results) were initially transmitted via fax or phone. The primary constraint on communication is the need to be HIPAA-compliant: for instance, faxes are, while ordinary (unencrypted) email is not. Even when existing systems transmit data to public health systems or other healthcare facilities, there continue to be issues with standardized data encoding so that the receiving group can process the data. There are many ways to have a more efficient, suitably secure system for communicating healthcare data, but also many challenges, primarily practical, to implementing such a system. At the same time, there are clear potential benefits to enabling rapid, secure communication of healthcare data, even within a single healthcare network. And of course, the benefits would be even greater when we consider the need for data integration and sharing between networks.

## 3.4. Data integration, sharing, and decentralization

The pandemic amplified issues healthcare has experienced in peacetimes in terms of integrating data streams and sharing data amongst healthcare and public health institutions. During times of crisis, data integration and interoperability becomes even more critical as diverse data sources–such as hospital records, public health reports, economic, and other indicators (e.g., medication and PPE purchase)–must be merged to provide a comprehensive understanding of the crisis and how to deploy appropriate interventions to different groups. The research community must not only determine which data sets should be included to facilitate timely decision-making for public health responses, but they also must provide *all* stakeholders with the understanding of how data is being used and possible implications of this data sharing.

There is a continued need to standardize reporting across healthcare institutions and public health sectors to ensure that data is shared and understood uniformly. We need protocols to verify the accuracy, granularity, and trustworthiness of data streams so that we can add novel data streams (e.g., wastewater tracking, travel patterns, household census data, purchasing habits, medication sales, school absences) to models and discontinue others that are not benefitting models. An open computing challenge continues to be how to deal with limited or small data sets in larger models. Although more and novel data streams could improve models and intervention plans, we must continue to empower individuals and communities to understand how their data is managed[7] and the risk-reward benefit of sharing data. From an institutional perspective, we need to create incentive models for healthcare systems to collaborate during surges and challenging times. Game theory research may be especially promising for designing incentive systems for people, communities, and organizations in ways that promote broader societal goals and benefits while maintaining the important autonomy of those individuals and groups.

Workshop attendees debated whether decentralized data would empower people and communities to better negotiate sharing of their data or further fragment data, which is challenging to merge during times of crisis. A continued struggle is how public health agencies access data; we need a meta-level of data access that continues privacy-preserving requirements (e.g., HIPAA), while allowing sharing of important data that impacts community health. We need to develop tools for getting and sharing genomic sequence pathogen data between academic and public health departments.

Another concern was equitable access to contributing and using data. Underserved communities are often not adequately represented in models, however the research community needs to investigate the socio-historical issues and incentive models for data contribution. In addition, smaller healthcare or public health institutions may not have the resources to gain access to the data sharing systems. One fruitful future pursuit could be creating an ecosystem of meta-level electronic health records (EHR) and personal health records

(PHRs) that focus on integration, data structures, collaboration, language translation, navigation, and patient experience. More work needs to be done to investigate the legal landscape of the health and public health related market identifying monopolies and data sharing–especially during times of crisis.

## 3.5. Strengthening communities

These data streams as described above can be used in decision support systems that balance between automation, people's input, and the burden of data (e.g., input, cleaning) to guide policies and guidelines. Attendees brainstormed creating predictive models to assess the impact of a disease outbreak on community and individual populations, and to use forecasting to inform decision-making. Ideally, these systems could create community-level early warning systems to provide information on the likelihood of hospital admissions, outpatient cases, and community-level impacts. These systems could be used in peacetime to anticipate the impact during flu season or heat waves.

The decision support systems should be built with micro- and macro-level models for analysis that could provide people with the ability to answer questions at a community, sub-population (e.g., schools, long term care), or individuals level over time. More work would be needed to identify how to get community-level and individual level input to help improve forecasting and create a responsive feedback loop. Attendees envisioned a system that could assist with tailoring public health interventions to the specific needs of sub-communities (e.g., county, city) level based on risk tolerance levels, understanding a community's past, current, and future responses to public health recommendations (e.g., a community resistant to "expert" medical recommendations, but trusting of local community group suggestions). The models could be expanded to assess the economic impact of interventions and how to assess the return on investment (e.g., school closures). The models could also help identify "behavioral cohorts" in terms of what motivates people to "buy in" to an intervention and then identify appropriate interventions based on similar groups.

---

[7] https://cra.org/ccc/wp-content/uploads/sites/2/2020/11/Modernizing-Data-Control_-Making-Personal-Digital-Data-Mutually-Beneficial-for-Citizens-and-Industry.pdf

Overall, more resources are needed in order to increase leadership, policy, and community communication during peacetime so that decision makers can build up social capital to use during challenging times. Currently, there is a need to communicate policy decisions and implementation strategies at all levels, and effectively communicate how communities can participate in decision-making. We also have to identify "trigger metrics" for when and how interventions will be deployed so that people can assess these metrics and reallocate resources as needed.

### 3.6. Strengthening healthcare

Throughout the country, but especially during pandemic times, Americans have experienced HCP shortages. We need to develop registry technologies to easily identify where HCP are, what knowledge they have, possibilities for retraining, and how sub-specialists can share relevant knowledge. An example of this was during the HIV and AIDS epidemic in the 1980s and 1990s, some specialists had "dial a doc" where healthcare providers around the country could call the number and talk to a specialist about how to care for a patient who had AIDS. Another promising area is to identify what current specialists can be retrained for other areas (e.g., anesthesiologists being redeployed during the pandemic in acute care), help healthcare communities plan for these redistributions, and provide materials for retraining and updating in an easily accessible format. Healthcare systems should identify resources needed to support HCP and identify appropriate incentive and compensation models to support them. Several nations have successfully deployed HCP registries and use them to manage their workforce [8]. Perhaps the U.S. can learn from these experiences and deploy a multi-tiered system that enables states and federal agencies to coordinate support for the nation's HCP and public health workers.

Supporting HCP during the COVID-19 pandemic came in many forms–from providing information to them as they need it in a readable format (especially when transferring patients between healthcare providers) to ensuring they have access to equipment they need. Current systems require burdensome manual input mechanisms (e.g., scanning each box of PPE) to locate, redistribute, and share equipment. Workshop attendees acknowledged the tradeoffs of understanding where equipment is located, but also emphasized the need to ensure equitable distribution of resources based on community conditions. Another research area to consider was how more sustainable practices could be integrated into the healthcare supply chain (e.g., extending shelf life; reducing waste). An example during the pandemic was how some vaccines had to be used in a certain time once opened, however there were challenges to communicate these vaccine openings to HCP effectively.

### 3.7. Strengthening the research enterprise

An area that could have significant impact in patient care and the advancement of science is iteratively defining, testing, and sharing best practices for patient treatment. The research community needs to have better mechanisms to collect innovative patient protocol data (e.g., testing accuracy, interpretation of results, treatment options and outcomes, a community's adherence to protocols) and develop benchmarks for when protocols should be disseminated for testing. Well-defined protocols for randomized control trials (RCTs) exist, however more work is needed to assess protocols within a shorter time period when illnesses are new or changing rapidly over time. More work is needed to develop benchmarks for conducting experiments and understanding how these real-world experiments translate to models for larger scale assessments. A sub-challenge here is standardizing data reporting on these iterative protocols as other institutions adopt these practices so that the research community can assess the protocol efficacy on a larger, more diverse dataset.

---

[8] Gilliam, Nora J., Dykki Settle, Luke Duncan, and Brian E. Dixon. "Chapter 14 - Health Worker Registries: Managing the Health Care Workforce." In Health Information Exchange (Second Edition), edited by Brian E. Dixon, 329-41: Academic Press, 2023.

Beyond collaboration between healthcare and public health institutions, more collaboration and coordination is needed among research institutions and experts. Workshop participants shared experiences about large projects between state health departments and multiple academic institutions that struggled to share data – thus easy to use, data sharing infrastructure is needed. One noticeable gap in these discussions was that most of the institutions were currently well resourced – thus, more attention should be given to ensure inclusive research with institutions that have varying resources. We recommend planning grants with low overhead to facilitate participation and collaboration between these research institutions with varying resources. Another area would be to have grant solicitations that emphasize and require diversity, equity, inclusion, and justice principles where researchers include underrepresented communities and international collaborations.

### 3.8. Peacetime foundation models to jumpstart pandemic models

A significant community research challenge lies in enabling actionable information in real time for HCP, healthcare systems, governments, schools, and other relevant entities. Such actionable information will prove especially useful during times of emergency (as it was the case at the beginning of the COVID-19 pandemic), however it can also be widely useful for peacetime healthcare situations. Multiple participants noted that there can be significant challenges with resource and staffing allocation, prediction, and monitoring even when not in the midst of a pandemic.

Foundation models (including, but not limited to, large language models) have proven to be quite powerful in terms of information aggregation, detection of patterns, and translation of low-level observations to actionable guidance[9]. Initial explorations have also demonstrated their potential for surfacing knowledge related to medical problems[10]. Moreover, they can typically be quickly fine-tuned for specific novel contexts or challenges. Especially in healthcare, specialized models could be developed to effectively account for different modalities of relevance, including language (e.g., EHR notes, patient-provider communication), vision (e.g., radiology images), behavioral signals (e.g., heart rate, blood analyses), and more–and thus are particularly appropriate for addressing these challenges in both peacetime and a pandemic. By building a peacetime foundation model and demonstrating the usefulness of this capability on a specific use case, such as a flu season, we can establish its practical value and potential impact. This same model can later be used as a starting point for emergency situations, such as emerging diseases or other unanticipated healthcare situations that require rapid solutions.

### 3.9. Data access and training

Much of the success of foundation models in a variety of domains has been attributed to their vast training data sets, which are drawn from a variety of online Web sources. Healthcare data is less readily available, which can potentially lead to model blind spots or accuracy issues. Current publicly and commercially available foundation models may not be directly applicable to the problems specific to the space of healthcare, and the size of such models may preclude individual healthcare networks from building their own.

We will need substantial data to develop healthcare-specific foundation models. Even with extensive patient records available across numerous hospitals, there are significant challenges associated with the use of such data. First, each patient's data, even if extensive, falls short of the vast volumes used to train current foundation models. This data sparsity is further exacerbated by missing patient information due to transitions between health systems. Second, there are often challenges associated with the data distribution, or with different hospitals and other data sources having different data characteristics that do not easily transfer across data repositories. Third, there are major challenges associated with the privacy of data, with many healthcare institutions being reluctant to make their data available even for internal model development, let alone sharing outside the institution.

---

[9] https://arxiv.org/abs/2108.07258

[10] https://arxiv.org/abs/2303.13375

We will also need to rethink model training. Predicting the next word in a sentence makes sense for language generation but will not immediately apply to health data, especially when presented in multimodal form (e.g., heart rate data paired with blood analyses). Such diversity of data is naturally occurring in healthcare, because of the numerous health data sources and the specialized knowledge and understanding necessary to make medical decisions. Significant advances have been made on multimodal foundation models, but many theoretical and practical challenges remain. Models will need instead to be trained by specifically accounting for "objectives" that are important in healthcare, such as improving diagnostic accuracy and patient outcomes; identifying and predicting health risks; personalizing treatment plans; optimizing resource allocation; facilitating clinical decision-making; and so on.

### 3.10. Infrastructure for data support

Leveraging the data available in different healthcare networks alongside powerful foundational models could significantly advance decision-making and other core processes in health. There are different strategies that can be used to connect data across networks, such as the QHIN framework[11]. One approach is to develop systems that can "translate" between data at different entities, without the data ever leaving its original location. Another alternative is a centralized data repository, potentially managed by the CDC, which could provide a unified platform for integrating diverse healthcare data. Key challenges will include merging heterogeneous datasets, determining the most relevant data elements, and incorporating novel data sources like hospital records, search queries, travel patterns, and school absences. Ensuring data privacy and establishing secure access protocols will also be crucial, as it will be exploring methods for incorporating public data from other countries to gain insights into global health trends. Establishing data trustworthiness criteria will also be essential for ensuring the reliability and accuracy of the repository. Finally, when direct data sharing is impractical, techniques like federated learning, where models are updated without centralizing the data, offer a viable alternative.

We need to develop base models that can handle the complexities of healthcare data to facilitate effective healthcare foundation models. These models should be able to address redundancy from different data sources, handle and fill in missing data, manage uncertainty, and identify correlations and relationships across the data. They should be accessible in a way that supports informed decision-making, while remaining explainable and transparent. Additionally, these models should be able to provide the necessary information for data visualizations and simulations. Clear policies need to be established to determine who can access the model and under what circumstances, with different access levels based on user needs and applications. In addition, we need more work in defining how models are validated and create transparency around model inputs and outputs. ONC's HTI-1[12] is a good first step to certify AI models that impact healthcare decisions, but we envision continued iteration will be needed on this process as AI continues to evolve.

## 4. Conclusion

Robust and timely computing research has the potential to better support HCPs to save lives in times of crisis (e.g., pandemics) and during "peacetime." The US healthcare system now is in dire need of systemic changes to enable HCP to provide high quality care and keep communities safe. In particular, models, data, and infrastructure all must be improved–including research, development, and implementation–to yield significant improvements in our healthcare system. This report has aimed to outline some of the key efforts and foci that could make a significant difference in both normal and crisis operations. HCP in the United States and abroad worked tirelessly throughout the pandemic, oftentimes with scarce resources, to help their neighbors. Research into how to make their jobs more efficient, impactful, and safe, is key to both preparing for the next pandemic, and making communities healthier today.

---

[11] https://rce.sequoiaproject.org/qhin-technical-framework-feedback-comments/

[12] https://www.healthit.gov/topic/laws-regulation-and-policy/health-data-technology-and-interoperability-certification-program

FUTURE OF PANDEMIC PREVENTION AND RESPONSE CCC WORKSHOP REPORT

## Appendix 1: Mass General Brigham Table Top Exercise After Action Report

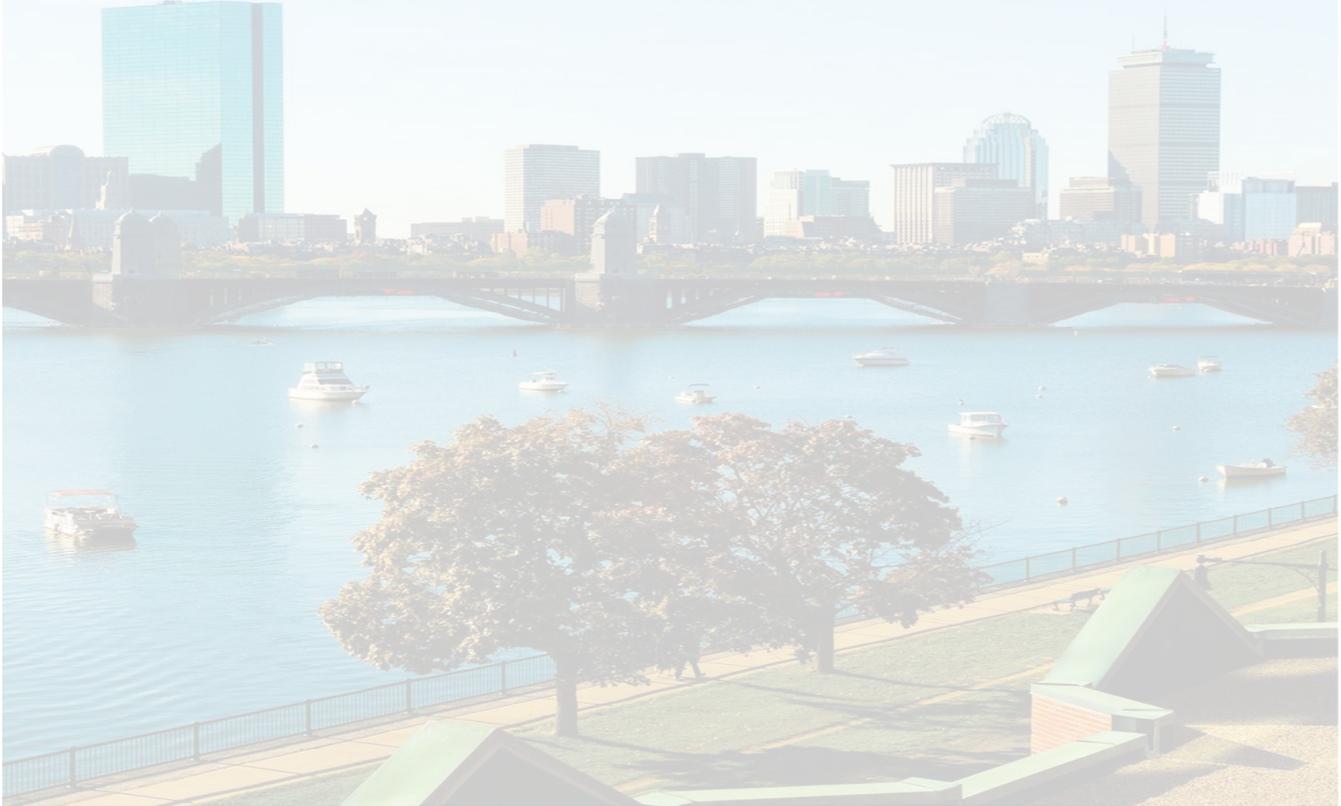

# Computing Community Consortium: Future of Pandemic Prevention and Response Exercise

September 20, 2023

Ann Arbor, MI

Executive Summary

## Executive Summary

The transdisciplinary nature of the Computing Community Consortium (CCC) allows for the application of high-impact research that strives to address the complex challenges our national and global communities face. As part of the CCC Future of Pandemic Response and Prevention Workshop on September 20-21, 2023, leaders from across disciplines and sectors came together to participate in a fictional exercise aimed at increasing understanding of how computing research and technology can be leveraged during future pandemic events to decrease morbidity and mortality.

The exercise was broken into the five modules listed below which challenged participants to explore novel and innovative solutions that could be created and mobilized in partnership across sectors.

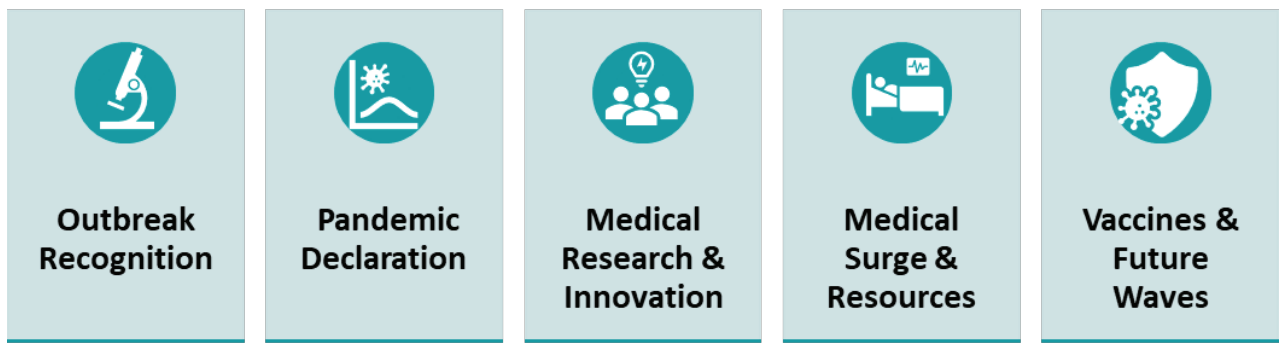

The exercise illustrated the potential value of the computing technology community in preparing for and responding to future pandemics and highlighted multiple untapped resources and capabilities that exist but require sustained development, coordination, and investment to codify into future use.

Several key themes emerged from the exercise discussions, including:

1. There are opportunities to advance structured mining of existing data sources and to optimize data sharing across platforms to enhance situational awareness during a pandemic.

2. Leveraging genomic data and sequencing may streamline pathways that can accelerate early diagnostics and testing during a pandemic.

3. Additional focused efforts are needed to explore novel technologies that can aid in the synthesis of medical and public literature and creation of community-based resources (e.g., decision support tools).

4. Developing flexible models that can be adjusted and updated rapidly based on new scientific knowledge is it emerges may facilitate the improved strategic adjustments that will be required in response to future pandemics (e.g., effectiveness of mask wearing or social distancing).

These findings represent a selected summary from the robust discussion from exercise participants. Detailed findings and barriers to implementation are described below.

## Exercise Findings by Module

*The following action steps are based on the module discussions, which were focused on how the computing community can support efforts, leverage existing technologies, foster, and build relationships, and create innovative solutions to support all phases of pandemic preparedness and response.*

### Module 1: Initial outbreak recognition

The beginning of an outbreak of an infectious disease is a key moment of opportunity for the computing research community to meaningfully affect the course of the response. For public health and medical leaders, effective management of an emerging or re-emerging pathogen depends on early recognition that a major outbreak or even pandemic may be developing. Computing community resources can improve upon existing systems to recognize subtle signals in data, to aggregate data across disparate sources, and provide continual support for sophisticated data analysis across all communities. Actionable next steps and opportunities for the computing community to consider include:

1. Investigate existing data sources (e.g., cell phone SIM card data, Google and Microsoft search history, Open Table, Uber, CVS, school absences, public transportation usage) that have the potential to provide insights into early behavioral patterns related to illness and disease avoidance, and engage key stakeholders involved in the collection of this data ahead of time for informative baseline data and to determine access and efficacy during infectious disease outbreaks.
2. Leverage genomic data from the identified pathogen to look for markers of change that may better assess severity potential and support creation of early diagnostics.
3. Explore the possibility of using computing technology to better identify patterns in diagnostics among confirmed or suspected cases to enable early recognition of disease in the absence of a definitive or reliable disease-specific test.
4. Leverage relationships with research institutions and universities embedded in affected countries or communities when attempting to gather data/information or conduct research.
5. Investigate opportunities to standardize the collection and reporting of health data in the United States to streamline analysis of trends.

### Module 2: Monitoring transformation from outbreak to pandemic

The evolution of an infectious disease outbreak to pandemic is a process that has opportunities where computing researchers can be engaged and provide actionable information that can assist healthcare and public health responders who attempt to predict the course of the incident and improve their response. These opportunities include:

1. Explore modeling capabilities for supply chain issues based on travel restrictions and products being mass purchased by large health care systems to help mitigate the impact of these issues to smaller hospitals with fewer resources and address hoarding of resources.
2. When implementing public health interventions, data sources may underreport or not capture communities with internet access limitations. There are significant opportunities for public health institutions to use social media platform data to identify effective messaging pathways and engage with community leaders, as well as reputable social media influencers, to assist with messaging/communication to priority populations before, during, and after public health emergencies (e.g., certificate programs, partnered communications).
3. Explore opportunities for artificial intelligence technology to augment existing mathematical modeling capabilities in developing artificial versions of impacted cities and populations to simulate various trends in disease spread and community impacts.
4. Develop flexible models that can be adjusted based on new scientific knowledge (e.g., effectiveness of mask wearing or social distancing). Include measures of public perception of reasonableness and effectiveness of non-pharmaceutical interventions.
5. Present information related to non-pharmaceutical interventions (e.g., outcome of adoption, outcome of ignoring) as a major part of public communications in response to provide up-to-date information, encourage individual health-seeking behavior, and relate to community needs.

## Module 3: Improving quality, inclusiveness, and utilization of medical research in the health response

The COVID-19 pandemic highlighted the disproportionate impact that outbreaks commonly have on vulnerable populations and communities of lower social economic status. COVID-19 also illustrated the immense challenge of collecting, vetting, and making available current research findings and best practices related to disease prevention and treatment. Opportunities for the computing community to support these efforts include:

1. Explore the use of artificial intelligence as a tool to enhance meta-analysis of emerging research such as narrowing scope, collating similar language, filtering articles with questionable study design, and identifying studies that contribute significant new scientific knowledge.
2. Build resources to encourage participation in clinical trials, including public-facing recruitment materials, outreach targeted to other health care facilities (e.g., urgent care, rural health centers), and decentralized systems for IRB and regulatory processes to enable more efficient and equitable research during infectious disease outbreaks/pandemics.
3. Create a widely accessible online platform that houses vetted materials and resources and fosters connections to experts (i.e., a call center).

## Module 4: Improving medical surge and management of scarce resources

The computing community is well positioned to help improve efforts to address the complex problems related to patient surge and resource scarcity during a pandemic. Opportunities to engage include:
1. Gather health data across disparate systems to make it sharable to increase trust and improve situational awareness. Establish defined, simple data metrics and allow organizations to establish bidirectional processes for sharing and reviewing data from others who submit. There are too many data sources, and they should be organized and consolidated.
2. Explore options for computing technology to help understand and influence care seeking behaviors and creation of decision-support tools.
3. Work with emergency medical services and hospitals to provide public/lawmakers with general awareness on factors such as impact to hospitals and available beds.

### Module 5: Mass distribution of medical countermeasures and future planning

Computing research can be leveraged to enhance information management, planning, and distribution of medical countermeasures. Examples of these opportunities include:

1. Partner with local community leaders to ensure public health messaging is communicated effectively and tailored to meet community needs.
2. Explore further opportunities to use genomic sequencing data and sources to enhance understanding of community disease prevalence and predict trends in the course of the pandemic (e.g., wastewater collection).
3. Utilize additional strategies beyond mortality metrics to use in targeting social distancing and other societal interventions (e.g., efforts to support return of children to classroom learning).
4. Develop integrated software solutions to improve understanding of and access to vaccine availability across providers that facilitates scheduling and accommodates walk-ins.

### Existing barriers to implementation
*In addition to the outlined recommendations, it is important to recognize the operational and logistical barriers in the public health and medical systems, as well as broader societal structures that may impact the ability to move this work forward. By addressing the following topic areas, computing leaders can help create a pathway for successful accomplishment of the calls to action established by this group.*

1. As was illustrated during the COVID-19 pandemic, public perception, engagement, and buy-in serve as a large driving force in the successful implementation of public health recommendations (e.g., masking, behavior change, vaccine uptake). Work is needed across sectors to assess how scientific and public health recommendations are shared to meet our communities where they are, and to begin to bridge the gap in areas of mistrust.
2. While many essential sources of data exist that can be accessed during infectious disease events, these systems and applications often do not have the ability to communicate with one another, resulting in very manual or duplicative processes when sharing data. Coupled with varying data needs (e.g., public health, healthcare, government) the inability to splice data in meaningful ways across sectors adds to the complexity of data sharing. The ability to mitigate these challenges would provide needed support during future pandemic scenarios.
3. Across sectors there are a diverse array of innovations, technologies, ideas, and resources that exist which are essential to future pandemic planning. Cross collaboration exists across some fields (e.g., academia, research, healthcare), but silos still exist. Failure to create and expand the ways in which we partner across fields will lead to continued impediments during future events.

## Conclusion

Accomplishing the above actions will require a concerted effort with sustained resources - including challenging the paradigm that exists in how our sectors currently interface and collaborate day-to-day. While it is unclear when the next infectious disease with pandemic potential will emerge, the work in front of us is essential to mitigate the barriers faced throughout the COVID-19 pandemic.

Innovations are taking place locally, nationally, and internationally to address this work, and the Computing Community Consortium is uniquely positioned to advocate, collaborate, and mobilize to support important next steps. There is great potential to leverage the diverse and creative ideas from this discussion to enhance pandemic preparedness in the future.

## Appendix 2: Workshop Participants

| First Name | Last Name | Company Name |
|---|---|---|
| YY | Ahn | Indiana University |
| Sidney | Allmendinger | Mass General Brigham |
| David | Banach | UConn Health |
| Andrew | Bartko | UC San Diego |
| Mike | Bell | CDC |
| Paul | Biddinger | Mass General Brigham |
| Westyn | Branch-Elliman | Harvard Medical School, VA National Artificial Intelligence |
| Tracy | Camp | Computing Research Association |
| Jennifer | Chien | UCSD |
| Ayush | Chopra | Massachusetts Institute of Technology |
| Theresa | Cullen | Pima County Public Health Department |
| Mary | Czerwinski | Microsoft Research |
| David | Danks | UC San Diego |
| Brian E. | Dixon | Regenstrief Institute-CBMI |
| Rob | Ernst | University of Michigan |
| Eleazar | Eskin | UCLA |
| Simon | Frost | Microsoft Premonition |
| Cat | Gill | CRA |
| Clint | Griffin | Emergency Physicians Inc./Holland Hospital emergency department |
| Haley | Griffin | CCC |
| Abba | Gumel | University of Maryland |
| Zaidat | Ibrahim | Indiana University Bloomington |
| Brendan | Kane | CRA |
| Andy | Kilianski | Advanced Research Projects Agency for Health (ARPA-H) |
| Larry | Madoff | Mass Department of Public Health |
| Madhav | Marathe | Biocomplexity institute at the University of Virginia |
| Juan Luis | Marquez | Washtenaw County Health Department |
| Rada | Mihalcea | University of Michigan |
| Emanuel | Moss | Intel Labs |
| Cody | Mullen | Purdue University |
| David | Reisman | Mass General Brigham |
| Alexander | Rodríguez | University of Michigan |
| Ann | Schwartz | CRA |
| Jennifer | Shearer | Massachusetts General Hospital |
| Erica | Shenoy | Mass General Brigham; Harvard Medical School |
| Katie | Siek | Indiana University |
| Mona | Singh | Princeton University |
| Michael | Sjoding | University of Michigan |
| Sarah | Tsay | Mass General Brigham |
| Anil | Vullikanti | University of Virginia |
| Sharon | Wright | Beth Israel Lahey Health |

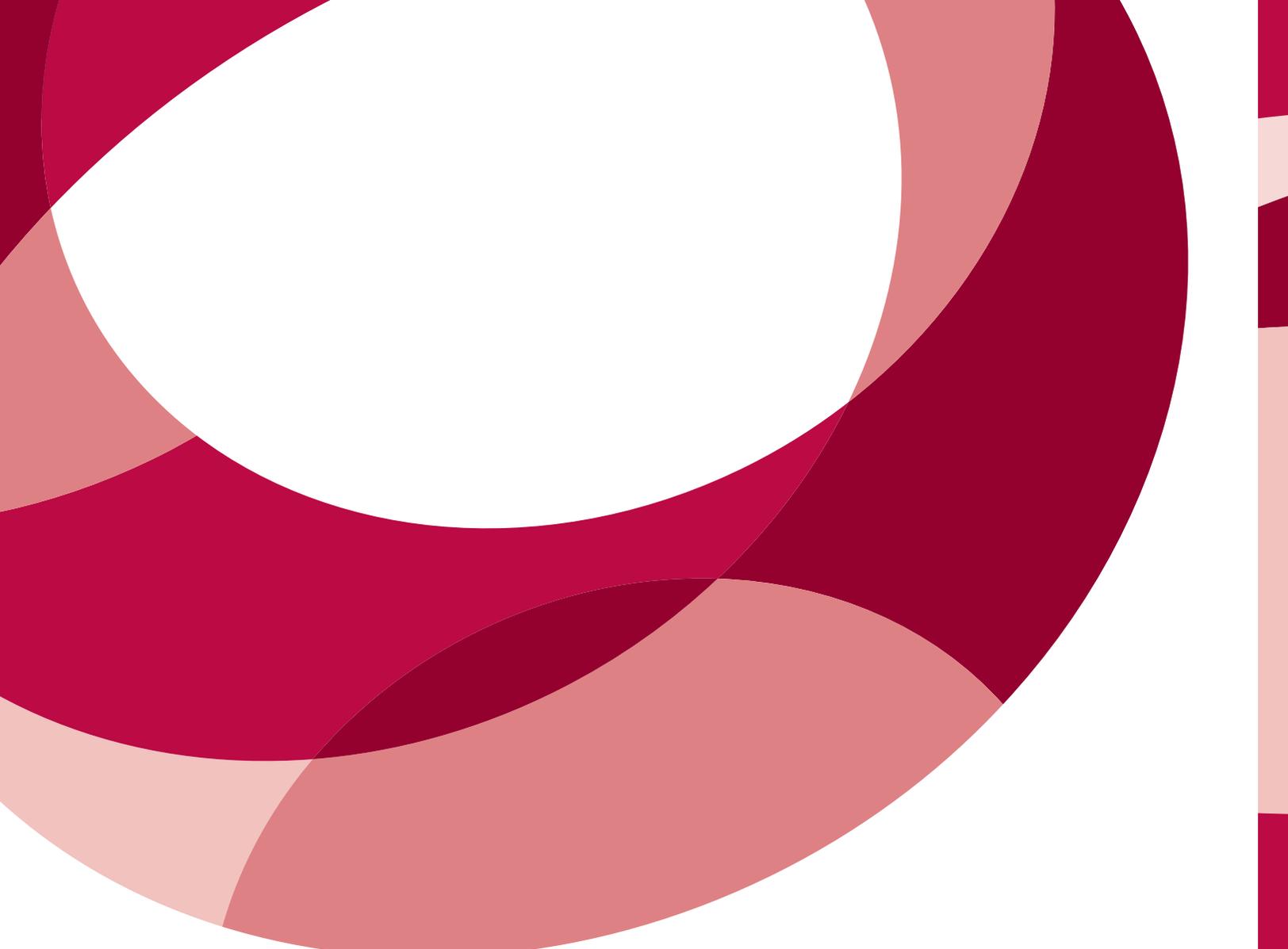

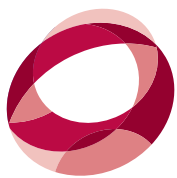

**CCC**
___
Computing Community Consortium
Catalyst

1828 L Street, NW, Suite 800
Washington, DC 20036
P: 202 234 2111 F: 202 667 1066
www.cra.org cccinfo@cra.org